\RequirePackage[pagewise,mathlines]{lineno}
\documentclass[twocolumn,aps,showpacs,superscriptaddress]{revtex4-1}
\usepackage{graphicx}% Include figure files
\usepackage{dcolumn}% Align table columns on decimal point
\usepackage{bm}% bold math
\usepackage{color}
\begin{document}
%\linenumbers
%\pagewiselinenumbers
%\rule{\textwidth}{0.3ex}\par
% A useful Journal macro
\def\Journal#1#2#3#4{{#1} {\bf #2}, #3 (#4)}

% Some useful journal names
\def\NCA{Nuovo Cimento}
\def\NIM{Nucl. Instr. Meth.}
\def\NIMA{{Nucl. Instr. Meth.} A}
\def\NPB{{Nucl. Phys.} B}
\def\NPA{{Nucl. Phys.} A}
\def\PLB{{Phys. Lett.}  B}
\def\PRL{Phys. Rev. Lett.}
\def\PRC{{Phys. Rev.} C}
\def\PRD{{Phys. Rev.} D}
\def\ZPC{{Z. Phys.} C}
\def\JPG{{J. Phys.} G}
\def\CPC{Comput. Phys. Commun.}
\def\EPJ{{Eur. Phys. J.} C}
\def\PR{Phys. Rept.}
\def\PRV{Phys. Rev.}
\def\JHEP{JHEP}

\preprint{}
\title{Energy dependence of acceptance-corrected dielectron excess mass spectrum at mid-rapidity in Au+Au collisions at $\sqrt{s_{NN}} = 19.6$ and 200 GeV}

\affiliation{AGH University of Science and Technology, Cracow 30-059, Poland}
\affiliation{Argonne National Laboratory, Argonne, Illinois 60439, USA}
\affiliation{Brookhaven National Laboratory, Upton, New York 11973, USA}
\affiliation{University of California, Berkeley, California 94720, USA}
\affiliation{University of California, Davis, California 95616, USA}
\affiliation{University of California, Los Angeles, California 90095, USA}
\affiliation{Universidade Estadual de Campinas, Sao Paulo 13131, Brazil}
\affiliation{Central China Normal University (HZNU), Wuhan 430079, China}
\affiliation{University of Illinois at Chicago, Chicago, Illinois 60607, USA}
\affiliation{Creighton University, Omaha, Nebraska 68178, USA}
\affiliation{Czech Technical University in Prague, FNSPE, Prague, 115 19, Czech Republic}
\affiliation{Nuclear Physics Institute AS CR, 250 68 \v{R}e\v{z}/Prague, Czech Republic}
\affiliation{Frankfurt Institute for Advanced Studies FIAS, Frankfurt 60438, Germany}
\affiliation{Institute of Physics, Bhubaneswar 751005, India}
\affiliation{Indian Institute of Technology, Mumbai 400076, India}
\affiliation{Indiana University, Bloomington, Indiana 47408, USA}
\affiliation{Alikhanov Institute for Theoretical and Experimental Physics, Moscow 117218, Russia}
\affiliation{University of Jammu, Jammu 180001, India}
\affiliation{Joint Institute for Nuclear Research, Dubna, 141 980, Russia}
\affiliation{Kent State University, Kent, Ohio 44242, USA}
\affiliation{University of Kentucky, Lexington, Kentucky, 40506-0055, USA}
\affiliation{Korea Institute of Science and Technology Information, Daejeon 305-701, Korea}
\affiliation{Institute of Modern Physics, Lanzhou 730000, China}
\affiliation{Lawrence Berkeley National Laboratory, Berkeley, California 94720, USA}
\affiliation{Massachusetts Institute of Technology, Cambridge, Massachusetts 02139-4307, USA}
\affiliation{Max-Planck-Institut fur Physik, Munich 80805, Germany}
\affiliation{Michigan State University, East Lansing, Michigan 48824, USA}
\affiliation{Moscow Engineering Physics Institute, Moscow 115409, Russia}
\affiliation{National Institute of Science Education and Research, Bhubaneswar 751005, India}
\affiliation{Ohio State University, Columbus, Ohio 43210, USA}
\affiliation{Institute of Nuclear Physics PAN, Cracow 31-342, Poland}
\affiliation{Panjab University, Chandigarh 160014, India}
\affiliation{Pennsylvania State University, University Park, Pennsylvania 16802, USA}
\affiliation{Institute of High Energy Physics, Protvino 142281, Russia}
\affiliation{Purdue University, West Lafayette, Indiana 47907, USA}
\affiliation{Pusan National University, Pusan 609735, Republic of Korea}
\affiliation{University of Rajasthan, Jaipur 302004, India}
\affiliation{Rice University, Houston, Texas 77251, USA}
\affiliation{University of Science and Technology of China, Hefei 230026, China}
\affiliation{Shandong University, Jinan, Shandong 250100, China}
\affiliation{Shanghai Institute of Applied Physics, Shanghai 201800, China}
\affiliation{Temple University, Philadelphia, Pennsylvania 19122, USA}
\affiliation{Texas A\&M University, College Station, Texas 77843, USA}
\affiliation{University of Texas, Austin, Texas 78712, USA}
\affiliation{University of Houston, Houston, Texas 77204, USA}
\affiliation{Tsinghua University, Beijing 100084, China}
\affiliation{United States Naval Academy, Annapolis, Maryland, 21402, USA}
\affiliation{Valparaiso University, Valparaiso, Indiana 46383, USA}
\affiliation{Variable Energy Cyclotron Centre, Kolkata 700064, India}
\affiliation{Warsaw University of Technology, Warsaw 00-661, Poland}
\affiliation{Wayne State University, Detroit, Michigan 48201, USA}
\affiliation{World Laboratory for Cosmology and Particle Physics (WLCAPP), Cairo 11571, Egypt}
\affiliation{Yale University, New Haven, Connecticut 06520, USA}
\affiliation{University of Zagreb, Zagreb, HR-10002, Croatia}

\author{L.~Adamczyk}\affiliation{AGH University of Science and Technology, Cracow 30-059, Poland}
\author{J.~K.~Adkins}\affiliation{University of Kentucky, Lexington, Kentucky, 40506-0055, USA}
\author{G.~Agakishiev}\affiliation{Joint Institute for Nuclear Research, Dubna, 141 980, Russia}
\author{M.~M.~Aggarwal}\affiliation{Panjab University, Chandigarh 160014, India}
\author{Z.~Ahammed}\affiliation{Variable Energy Cyclotron Centre, Kolkata 700064, India}
\author{I.~Alekseev}\affiliation{Alikhanov Institute for Theoretical and Experimental Physics, Moscow 117218, Russia}
\author{J.~Alford}\affiliation{Kent State University, Kent, Ohio 44242, USA}
\author{A.~Aparin}\affiliation{Joint Institute for Nuclear Research, Dubna, 141 980, Russia}
\author{D.~Arkhipkin}\affiliation{Brookhaven National Laboratory, Upton, New York 11973, USA}
\author{E.~C.~Aschenauer}\affiliation{Brookhaven National Laboratory, Upton, New York 11973, USA}
% ...

\collaboration{STAR Collaboration}\noaffiliation
\date{\today}
\begin{abstract}
The acceptance-corrected dielectron excess mass spectra, where the known hadronic sources have been subtracted from the inclusive dielectron mass spectra, are reported for the first time at mid-rapidity $|y_{ee}|<1$ in minimum-bias Au+Au collisions at $\sqrt{s_{NN}}$ = 19.6 and 200 GeV. The excess mass spectra are consistently described by a model calculation with a broadened $\rho$ spectral function for $M_{ee}<1.1$ GeV/$c^{2}$. The integrated dielectron excess yield at $\sqrt{s_{NN}}$ = 19.6 GeV for $0.4<M_{ee}<0.75$ GeV/$c^2$, normalized to the charged particle multiplicity at mid-rapidity, has a value similar to that in In+In collisions at  $\sqrt{s_{NN}}$ = 17.3 GeV. For $\sqrt{s_{NN}}$ = 200 GeV,  the normalized excess yield in central collisions is higher than that at $\sqrt{s_{NN}}$ = 17.3 GeV and increases from peripheral to central collisions. These measurements indicate that the lifetime of the hot, dense medium created in central Au+Au collisions at $\sqrt{s_{NN}}$ = 200 GeV is longer than those in peripheral collisions and at lower energies.
\end{abstract}

\maketitle

%\begin{keyword}
%dielectron excess spectrum, chiral symmetry restoration, vector meson in-medium modifications, %lifetime of hot, dense medium


%\end{keyword}

%\end{frontmatter}


\section{Introduction}
\label{introduction}

Dileptons are crucial probes for studying the properties of the strongly interacting, hot and dense matter which is created in ultrarelativistic heavy-ion collisions at the Relativistic Heavy-Ion Collider (RHIC)~\cite{qgp1,qgp2}. They are produced during the whole evolution of the created matter, and are not subject to strong interactions with the medium. Dielectron pairs are sensitive probes of the medium properties throughout the spacetime evolution of the medium~\cite{probe1,probe2} because they are produced through a variety of mechanisms and in several different kinematic regimes.

In the low invariant mass region, $M_{ll}<1.1$ GeV/$c^{2}$ (LMR), the dilepton production is dominated by in-medium decay of vector mesons ($\rho$, $\omega$ and $\phi$) in the hadronic gas phase. In-medium modifications to the mass and width of the vector mesons are considered a link to chiral symmetry restoration~\cite{probe1,probe2}. In the vacuum, chiral symmetry is spontaneously broken, which results in mass differences between chiral partners [e.g. $\rho$ and $a_{1}(1260)$]. In the hot, dense medium, chiral symmetry is expected to restore and the mass distributions of $\rho$ and $a_{1}(1260)$ are expected to change and degenerate. Since it is extremely challenging to measure a spectral function for the $a_{1}(1260)$ meson, one cannot directly observe the disappearance of the mass splitting between the $\rho$ and $a_{1}(1260)$ experimentally. Instead, efforts are devoted to studying the modification of vector meson spectral function. Two schematic scenarios are used to describe the in-medium $\rho$ spectrum function: a broadened and a dropping-mass $\rho$. The broadened $\rho$ scenario incorporates finite temperature effects into self-energy corrections through medium interactions and $\pi\pi$ annihilations~\cite{broaden}. The dropping mass scenario uses the quark mean field from a high temperature/density regime wherein constituent quarks are the relevant degrees of freedom, and then extrapolates down to a low temperature/density regime wherein hadrons are appropriate degrees of freedom~\cite{dropping}. 

The CERES experiment at the CERN-SPS reported an excess dielectron yield with respect to the known hadronic sources in the LMR in Pb+Au collisions at $\sqrt{s_{NN}}$ = 17.2 GeV, which indicates that the vector mesons are modified in medium~\cite{CERES}. More recently, NA60 published a precise measurement of the dimuon invariant mass spectra in In+In collisions at $\sqrt{s_{NN}}$ = 17.3 GeV~\cite{NA60}. The results show a significant excess in the LMR above the hadronic sources. In both cases, the excess is consistent with a broadened $\rho$ spectral function~\cite{broaden}, but not with a $\rho$ dropping-mass scenario~\cite{dropping}, where both models have been evaluated for the same fireball evolution. In the model calculation, the coupling to the baryons in the medium plays a dominant role in the broadening of the $\rho$ spectral function~\cite{broaden,CERES,NA60}.

At RHIC, a significant enhancement in the dielectron continuum, compared with the known hadronic sources, has been observed in the LMR by both the PHENIX and STAR collaborations in Au+Au collisions at $\sqrt{s_{_{NN}}} = 200$ GeV~\cite{eePHENIXaa200,eeSTARaa200}. Results from the STAR collaboration show that the excess dielectron yield in the mass region 0.3-0.76 GeV/$c^2$ follows an $N_{part}^{1.54\pm0.18}$ dependence, where $N_{part}$ is the number of participant nucleons in a collision~\cite{eeSTARaa200}. However, the PHENIX Collaboration reported  significant higher excess dielectron yields in central collisions~\cite{eePHENIXaa200}. Theoretical calculations~\cite{ralf, ralf:08, PHSD:12, USTC:12}, which describe the SPS dilepton data, fail to consistently describe the low-mass enhancement at low transverse momentum ($p_T$) observed by PHENIX in both 0-10\% and 10-20\% central Au+Au collisions~\cite{eePHENIXaa200}. The same calculations, however, correctly describe the STAR measurement of the low-$p_T$ and low-mass enhancement from peripheral to central Au+Au collisions~\cite{eeSTARaa200}. While the discrepancy between STAR and PHENIX in central Au+Au collisions at $\sqrt{s_{_{NN}}} = 200$ GeV is still under investigation, it is important to have dilepton measurements at RHIC at lower beam energies with the same large acceptance as for the 200 GeV data. Since the total baryon density does not change significantly from $\sqrt{s_{NN}} = $ 17.3 GeV to $\sqrt{s_{NN}} = $  200 GeV~\cite{BESwhiteII}, it is essential to confirm that the broadened $\rho$ spectral function, which describes the results at 17.3 GeV and the 200 GeV STAR data, is consistent with the 19.6 GeV results.

In the intermediate mass region, $1.1<M_{ll}<3.0$ GeV/$c^{2}$ (IMR), dilepton production is expected to be directly related to thermal radiation of the partonic phase, which is considered to be the prime signature of deconfinement~\cite{ralf,ralf:08}. An enhanced yield in this region was first observed by HELIOS/3~\cite{HELIOS3} and NA38/NA50~\cite{NA3850}. More recently, the NA60 collaboration reported an enhancement in the IMR which cannot be connected to decays of \textit{D} mesons, but may be the result of thermal radiation~\cite{NA60}. However, it is experimentally challenging to extract the signal in the presence of significant background sources from open heavy-flavor semi-leptonic decays, such as $c\bar{c}\rightarrow l^{+}l^{-}X$ or $b\bar{b}\rightarrow l^{+}l^{-}X$. 

In this letter, we report the first dielectron measurements at mid-rapidity in minimum-bias Au+Au collisions at $\sqrt{s_{NN}}$ = 19.6 GeV with the STAR detector~\cite{STAR}.  Furthermore, we present the first acceptance-corrected dielectron excess mass spectra in Au+Au collisions at $\sqrt{s_{NN}} = $ 19.6 and 200 GeV which are compared with measurements from NA60 and theoretical model calculations. 
The invariant excess dielectron spectra at different centralities and energies allow for a first systematic study of the lifetime of the hot, dense medium using electromagnetic probes at RHIC. It was pointed out that the excess dielectron yield at low mass is proportional to the total lifetime of the hot, dense medium at $\sqrt{s_{NN}}$ = 6-200 GeV~\cite{rapp:14}.

\section{Experiment and data analysis}
\label{analysis}
In this analysis, 33 million minimum-bias (MB) Au+Au  (0-80\%) events at $\sqrt{s_{NN}} = $ 19.6 GeV, recorded by the STAR experiment in the year 2011, were used. The results at $\sqrt{s_{NN}} = $ 200 GeV are derived from the same data analysis reported in Ref.~\cite{eeSTARaa200}. The MB trigger at $\sqrt{s_{NN}} = $ 19.6 GeV was defined as a coincidence of the two Beam Beam Counters covering the pseudorapidity range $3.3\!<\!|\eta|\!<\!5.0$~\cite{BBC}. Charged tracks were reconstructed by the Time Projection Chamber (TPC)~\cite{TPC}, which has full azimuthal coverage at $|\eta|<1$. The absolute distance between collision vertices and the TPC center along the beam direction was required to be less than 70 cm. The transverse momentum resolution is measured to be $\Delta p_{T}/p_{T}=0.01\times[1+p_{T}/(2$ GeV/$c)]$ for $p_T\!<$ 5 GeV/$c$. The Time-Of-Flight (TOF)~\cite{TOF} detector, which covers the pseudorapidity range $|\eta|<0.9$, provides the arrival time of charged tracks from the collision vertex. Slow hadrons can be rejected by a velocity cut $|1/\beta-1/\beta_{\rm exp}|<0.025$ in the range of $0.2<p_{T}<3$ GeV/$\textit{c}$, where $\beta$ is the measured velocity and $\beta_{exp}$ is the expected velocity calculated using the track length and momentum with the assumption of the electron mass. After the velocity cut, electron identification is achieved by cutting on the normalized ionization energy loss ($n\sigma_{e}=\log(\frac{dE}{dx}/I_{e})/R_{e}$) measured by the TPC, where $dE/dx$ is the energy loss, $I_{e}$ is the expected $dE/dx$ for an electron and $R_{e}$ is the $dE/dx$ resolution of an electron, which is better than 8\%~\cite{Bichsel}. 
The $n\sigma_{e}$ cut is momentum dependent and results in a high electron purity of $>$ 93\% and an efficiency of $>$ 65\% on average~\cite{eeSTARaa200,pp200ee}. 

\begin{figure}[htbp]
\begin{center}
\includegraphics[width=0.48\textwidth]{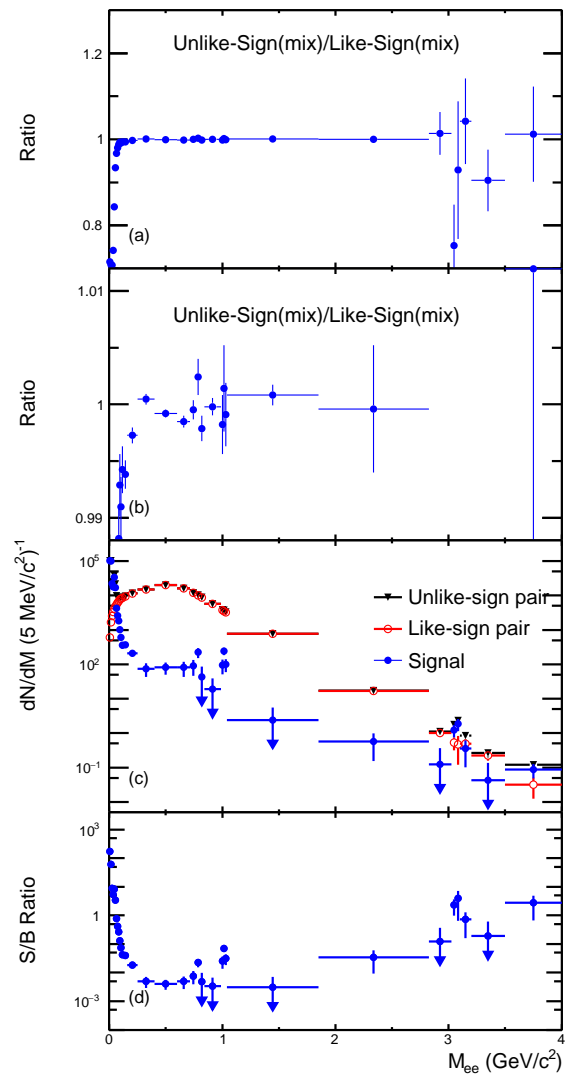}
\caption{(color online) (a): Ratio of mixed-event unlike-sign pair to mixed-event like-sign pair dielectron mass distributions. (b) A zoom-in version of Panel (a). (c): Reconstructed dielectron unlike-sign pairs (inverted triangles), like-sign pairs (open circles) and signal (filled circles) distributions. (d): The signal to background ratio ($S/B$). All panels are presented as a function of dielectron invariant mass in Au+Au collisions at $\sqrt{s_{NN}}$ = 19.6 GeV.}\label{fig:eemass} 
\end{center}
\end{figure}

The electron and positron candidates are paired by opposite and same sign charges, called unlike-sign and like-sign pairs, respectively. The like-sign pairs are used to statistically reproduce the combinatorial and correlated pair backgrounds. The combinatorial background comes from two random tracks without correlation. The correlated background is the result of two electrons, each of which comes from a different but correlated process of a particle decay or a jet fragmentation. For example, consider a $\pi^{0}\rightarrow \gamma e^{+}e^{-}$ Dalitz decay where the gamma may convert on some material to form an additional  $e^{+}e^{-}$ pair. The $e^{\pm}$ from the $\pi^{0}$ paired with a $e^{\mp}$ from the $\gamma$ can produce a correlated background pair. This correlated background can be reproduced by like-sign pairs.

The unlike-sign and like-sign pairs have different acceptances due to dead areas of the detector and the different bending curvatures of positively and negatively charged particles in the magnetic field. The dead area fraction is 13\% along the azimuthal distribution at $\eta\!<\!1$. A mixed-event technique~\cite{eePHENIXaa200} is applied to estimate the acceptance differences between the unlike-sign and like-sign distributions. Figure~\ref{fig:eemass} (a) shows the ratio between mixed-event unlike-sign pairs and mixed-event like-sign pairs as a function of dielectron mass. A zoom-in version is shown in Fig.~\ref{fig:eemass} (b).

The background subtraction is based on the measured like-sign spectra with the assumption that the shape and magnitude of the correlated background are the same in the unlike- and like-sign spectra. We subtract the like-sign background (corrected for the acceptance difference using the mixed event technique mentioned above) from the unlike-sign distributions to obtain the raw dielectron signals. The mixed-event background is not used for background subtraction, since the correlated background contribution is difficult to address with limited statistics at $M_{ee}>1.5$ GeV/$c^{2}$ for $\sqrt{s_{NN}}$ =19.6 GeV. Figure~\ref{fig:eemass} (c) shows the invariant mass distributions of unlike-sign pairs, like-sign pairs and background-subtracted signals. The signal to background ratio is shown in Fig.~\ref{fig:eemass} (d). Dielectron pairs from photon conversions in the detector materials are suppressed by selecting tracks with a distance of closest approach to the collision vertex that is less than 1 cm,  and a minimum opening angle cut between the two electron candidates~\cite{eePHENIXaa200,eeSTARaa200}. The minimum opening angle is 0.84 rad at $M_{ee}\!<\!0.03$ GeV/$c^2$ and decreases as a function of $M_{ee}$ according to a function form of $A/[B+exp(C/M_{ee})]$, in which A, B, and C are input parameters. For $M_{ee}\!>\!0.1$ GeV/$c^2$, the minimum opening angle is zero.

\begin{figure}[hbtp]
\centering
\includegraphics[width=.5\textwidth]{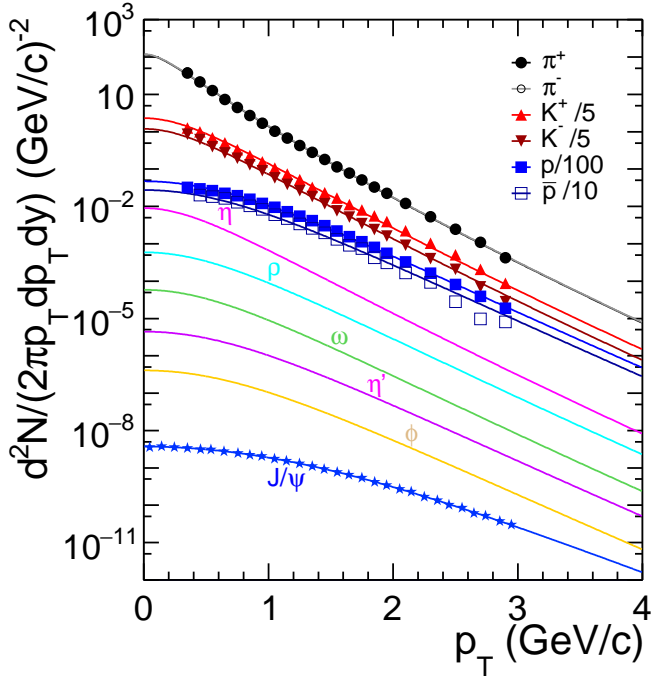}
\caption{(color online) The Tsallis Blast Wave (TBW) function fit~\cite{Tsallis1, Tsallis2} to the NA49 $p_T$ spectra of pions, kaons and protons in Pb+Pb at $\sqrt{s_{NN}}$ = 17.3 GeV~\cite{NA49}. The data points o f $\pi^{+}$ completely overlap with that of $\pi^{-}$ on the figure. Other meson $p_{T}$ spectra are predicted by the TBW function. For $J/\psi$, the $p_T$ shape is determined by an independent TBW function fit to the $J/\psi$ $p_T$ spectra measured by NA50~\cite{NA50jpsi}. More details are in the text.}\label{fig:tsallisfit} 
\end{figure}

\begin{figure}[hbtp]
\centering
\includegraphics[width=.5\textwidth]{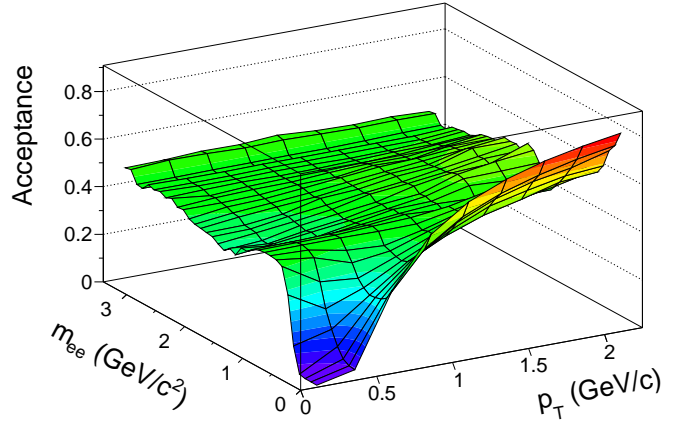}
\caption{(color online) The acceptance of virtual photon decayed dielectrons in the STAR detector in Au+Au collisions at $\sqrt{s_{NN}} = 19.6$ GeV.}\label{fig:acceptance} 
\end{figure}

\begin{figure}[hbtp]
\centering
\includegraphics[width=.5\textwidth]{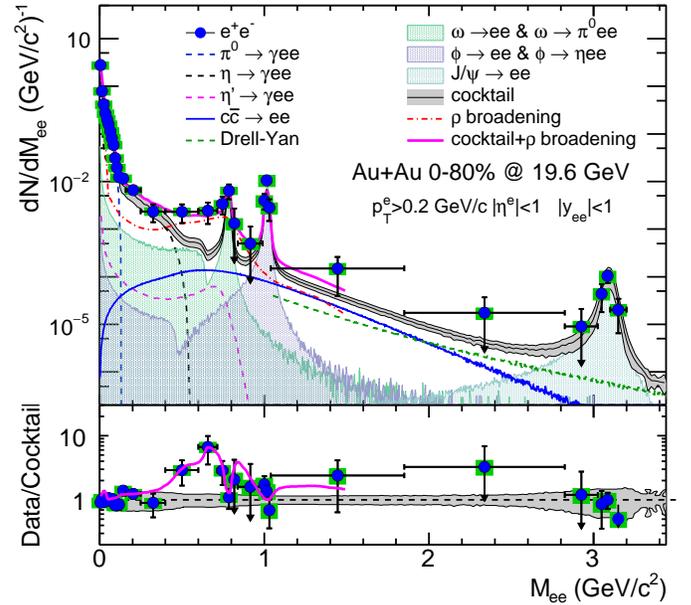}
\caption{(color online) Dielectron invariant mass spectrum in the STAR acceptance ($|y_{ee}|<1$, $0.2<p^{e}_{T}<3$ GeV/$c$, $|\eta^{e}|<1$) after efficiency correction, compared with the hadronic cocktail consisting of the decays of light hadrons and correlated decays of charm in Au+Au collisions at $\sqrt{s_{NN}} = 19.6$ GeV. The data to cocktail ratio is shown in the bottom panel. Theoretical calculations~\cite{ralf,ralfPrivate} of a broadened $\rho$ spectral function are shown up to 1.5 GeV/$c^{2}$ for comparison. Systematic uncertainties for the data points are shown as green boxes, and the grey band represents the uncertainties for the cocktail simulation.}\label{fig:eemasscor} 
\end{figure}

The raw dielectron signal is corrected for the electron reconstruction efficiency. The single electron reconstruction efficiency includes TPC tracking, electron identification and TOF matching efficiencies. The TPC tracking efficiency is determined by embedding Monte Carlo (MC) tracks into real raw data events, processing the track reconstruction with a GEANT model of the STAR detector~\cite{GEANT}, and determining the fraction of those embedded MC tracks which are reconstructed as good tracks. The efficiency correction includes the effect of dead areas in the detector.
The TOF matching and electron identification efficiencies are reproduced from real data. Detailed procedures to obtain the TPC and TOF efficiencies are explained in Ref.~\cite{pp200ee}. The energy loss and bremsstrahlung radiation effects for electrons are reproduced by the GEANT simulation. The single electron efficiency is convoluted into the pair efficiency with the decay kinematics in the simulation. 

The hadronic sources of dielectron pairs include: Dalitz decays $\pi^{0}\rightarrow \gamma e^{+}e^{-}$, $\eta \rightarrow \gamma e^{+}e^{-}$ and $\eta^{\prime}\rightarrow \gamma e^{+}e^{-}$; vector meson decays: $\omega \rightarrow \pi^{0} e^{+}e^{-}$, $\omega \rightarrow e^{+}e^{-}$, $\rho^{0} \rightarrow e^{+}e^{-}$, $\phi \rightarrow \eta e^{+}e^{-}$, $\phi \rightarrow e^{+}e^{-}$ and $J/\psi \rightarrow e^{+}e^{-}$; heavy-flavor hadron semi-leptonic decays: $c\bar{c} \rightarrow e^{+}e^{-}X$; Drell-Yan. The $\rho$ meson contribution is not evaluated in the simulation, but included in the model calculation (as described in Sec. III). The $b\bar{b} \rightarrow e^{+}e^{-}X$ process is not included as it has negligible contribution to the cocktail in Au+Au collisions at $\sqrt{s_{NN}}$ = 19.6 GeV. 

The input hadron spectra to the cocktail are derived from a Tsallis Blast Wave (TBW) function fit~\cite{Tsallis1, Tsallis2} to the NA49 $p_T$ spectra of pions, kaons and protons in Pb+Pb at $\sqrt{s_{NN}}$ = 17.3 GeV~\cite{NA49}, as shown in Fig.~\ref{fig:tsallisfit}. Other meson $p_{T}$ spectra are predicted by the TBW function using the same freeze-out parameters from $p_{T}$ fit of pions, kaons and protons. The extra uncertainty caused by the input $p_{T}$ spectra is found to be less than 10\% and has been propagated to the final cocktail uncertainty. For $J/\psi$, the $p_T$ shape is determined by an independent TBW function fit to the $J/\psi$ $p_T$ spectra measured by NA50~\cite{NA50jpsi}.

The $\pi^{0}$ contribution is obtained by matching the dielectron mass distribution from simulated $\pi^{0}\rightarrow \gamma e^{+}e^{-}$ and $\eta\rightarrow \gamma e^{+}e^{-}$ decays to the efficiency-corrected dielectron mass spectrum for $M_{ee}\!<0.1$ GeV/$c^2$. We also match the $J/\psi\rightarrow e^{+}e^{-}$ distribution from simulation to the measured dielectron production in the corresponding mass region. The meson yields of other mesons are derived by the meson-to-pion  ratios~\cite{CERES} and the pion yields. Table~\ref{tab:yield} lists the integrated yields used in the simulation at mid-rapidity for Au+Au collisions at $\sqrt{s_{NN}} = 19.6$ GeV. The branching ratios of mesons to dielectrons and their uncertainties are from Ref.~\cite{pdg2012}.

\begin{table}\caption{The meson yields, $dN/dy$, at mid-rapidity used in the hadronic cocktail for 0-80\% Au+Au collisions at $\sqrt{s_{NN}}$ = 19.6 GeV. The uncertainty includes contributions from the TBW fit and the meson-to-pion ratio.  \label{tab:yield}} \centering
\vspace{0.1in}
\begin{tabular}{c|c|c} 
\hline\hline
 meson yield & $dN/dy$  & uncertainty (\%) \\
\hline
$\pi^{0}$   &  49.6  & 8 \\
$\eta$   &  4.22 & 14 \\
$\omega$ & 3.42 & 16 \\
$\phi$   & 0.89 & 13 \\
$\eta^{\prime}$ & 0.39 & 17 \\
$J/\psi$   & $2.18\times10^{-4}$ & 32 \\
\hline \hline
\end{tabular}
\end{table}

The $e^{+}e^{-}$ mass distribution from open heavy-flavor sources is generated using PYTHIA 6.416~\cite{pythia}. Previous charm cross section measurements from the SPS, FNAL, STAR and PHENIX experiments ~\cite{charmdata} are well described by the upper limit of a Fixed-Order Next-to-Leading Logarithm (FONLL) calculation~\cite{FONLL:12}. Therefore we obtain the charm total cross section in $p+p$ at $\sqrt{s}$ = 19.6 GeV by scaling the FONLL upper limit to the previous measurements using the minimum $\chi^{2}$ method.  This total cross section $8.2\pm0.5$ $\mu$b is used to normalize the dielectron yield from the PYTHIA simulation, which is additionally scaled by the number of binary collisions for Au+Au at $\sqrt{s_{NN}}$ = 19.6 GeV to be compared with the data.

For the efficiency-corrected dielectron invariant mass distribution, the systematic errors are dominated by uncertainties on the TPC tracking efficiency (14\% in the dielectron yields), the TOF matching efficiency (10\% in the dielectron yields), hadron contamination (0-20\%), and electron identification (2\%). The total systematic uncertainty on the pair reconstruction efficiency is estimated to be 18\%. The systematic uncertainties on the like-sign background subtraction were mainly from the uncertainties on the acceptance difference factors between the unlike-sign and like-sign pairs. The acceptance difference factors were derived using mixed-event technique.
In the mixed-event technique, tracks from different events were used to form unlike-sign or like- sign pairs. The events were divided into different categories according to the collision vertex, event plane, azimuthal angle, and centrality. The bin sizes of collision vertex, event plane, azimuthal angle, and centrality were chosen to be small enough and the two events to be mixed must come from the same event category to ensure similar detector geometric acceptance, azimuthal anisotropy, and track multiplicities. The uncertainties in the acceptance difference factors were found to be 0.003\% and result in 1\% uncertainties for the dielectron signals.
For the cocktail simulation, the systematic uncertainties come from the uncertainties of particle yields, decay branching ratios and form factors. Table~\ref{tab:uncertainty} lists all the contributions to the systematic uncertainties on the dielectron mass spectrum and cocktail simulation within the STAR acceptance at $\sqrt{s_{NN}} = 19.6$ GeV.

After efficiency correction, the dielectron excess mass spectrum is corrected for the detector acceptance. The acceptance correction is estimated by a Monte Carlo simulation with inputs of virtual photon yield spectra, phase space distributions and decay kinematics. The method is similar to the approach used by NA60~\cite{NA60Acceptance}, in which one assumes that the excess yields are from medium emission. The acceptance is calculated by the yield ratio of reconstructed dielectrons in the STAR detector to the input dielectrons. Figure~\ref{fig:acceptance} shows the two-dimensional acceptance of the virtual photons with a Gaussian-like rapidity distribution in Au+Au at $\sqrt{s_{NN}} = 19.6$ GeV at STAR. The $\sigma$ value of the distribution is 1.5~\cite{NA60Acceptance}. The same approach was used in Au+Au at $\sqrt{s_{NN}} = 200$ GeV except that we used a flat rapidity distribution as our default case. The acceptance correction factor at $\sqrt{s_{NN}}$ = 200 GeV differs from that at $\sqrt{s_{NN}}$ = 19.6 GeV by 5\% mainly due to the input $p_T$ spectra of virtual photons.

For the dielectron excess mass spectrum, additional systematic uncertainties come from the subtraction of the cocktail contribution and the acceptance correction. In Au+Au at $\sqrt{s_{NN}} = 200$ GeV, the cocktail simulation is detailed in Ref.~\cite{eeSTARprc:15}. For the charm correlation contribution, we studied the following cases: a) keep the direct PYTHIA correlation between $c$ and $\bar{c}$ which was used in our default cocktail calculations; b) break the azimuthal angular correlation between charm decayed electrons completely but keep the $p_T$, $\eta$, and $\phi$ distributions from PYTHIA; c) randomly sample two electrons with the single electron $p_T$, $\eta$, and $\phi$ distributions from PYTHIA; and d) based on c), but sample the $p_T$ of each electron according to the modified $p_T$ distribution from the measurements of non-photonic electron nuclear modification factors in Au+Au collisions. The maximal difference between case a) and the other three is taken as the systematic uncertainties on the charm correlation contribution.

The uncertainty from acceptance correction contains uncertainties from the rapidity distribution and input dielectron sources. A uniform rapidity distribution is compared with the Gaussian-like case, and the resulting uncertainty is 2\% in the LMR in Au+Au at $\sqrt{s_{NN}} = 19.6$ GeV. For 200 GeV, we used a pion rapidity distribution to compare to the default case and quoted the difference between them as systematic uncertainty, which is about 2\%.
The uncertainty from the input $p_{T}$ spectrum is at the same level as the rapidity distribution uncertainty.

We also obtain the acceptance of the excess dielectrons from model calculations~\cite{ralfPrivate}. The difference between the simulation and theoretical calculation is about 20\% for $M_{ee}<0.4$ GeV/$c^{2}$ and less than 10\% for $M_{ee}>0.4$ GeV/$c^{2}$. It is included in the excess yield uncertainties.

%%%%%%%%%%%%%%%%%%%%%%%%%%%%%%%%%%%%%%%%%%%%%%%%%%%%%%%%%%%%%%%%%%%%%%

\section{Results and discussion}
\label{results}
The dielectron invariant mass distribution after efficiency correction is shown in the upper panel of Fig.~\ref{fig:eemasscor} for Au+Au collisions at $\sqrt{s_{NN}} = 19.6$ GeV. It is compared with a hadronic cocktail simulation, which consists of all the dielectron hadronic sources except the $\rho^{0}$. An enhancement of the dielectron yield is observed in the mass region $M_{ee}<1$ GeV/$c^{2}$. A model calculation with a broadened $\rho$ spectral function~\cite{ralf:08} is added to the hadronic cocktail and compared with the data,  as shown in the bottom panel of Fig.~\ref{fig:eemasscor}. The dielectron yields in the model calculation were filtered by the STAR acceptance ($p^{e}_{T}>0.2$ GeV/$c$ and $|\eta^{e}|<1$). The model calculation involves a realistic space-time evolution, and includes contributions from quark-gluon-plasma (QGP), 4-pion annihilation and in-medium vector meson contributions. The initial temperature from the model is 224 MeV and the starting time $\tau_{0}$ is ~0.8 fm/c~\cite{ralfPrivate}. The comparison of the model with data shows that a broadened $\rho$-spectra scenario is consistent with the STAR data within uncertainties. The same conclusion has been drawn in Au+Au collisions at $\sqrt{s_{NN}} = 200$ GeV~\cite{eeSTARaa200}. Using the broadened $\rho$ spectral function, QCD and Weinberg sum rules, and inputs from Lattice QCD, theorists have demonstrated that when the temperature reaches 170 MeV, the derived $a_{1}(1260)$ spectral function is the same as the in-medium $\rho$ spectral function, a signature of chiral symmetry restoration~\cite{Rapp:13}. 

\begin{table}[t]\caption{Summary of systematic uncertainties for the measured dielectron mass spectrum and simulated cocktail within STAR detector acceptance in Au+Au at $\sqrt{s_{NN}} =$ 19.6 GeV. The uncertainty on hadron contamination leads to a mass-dependent uncertainty for the measured dielectron continuum. The uncertainties of particle yields, branching ratios, and form factors result in mass-dependent uncertainties for the simulated cocktail. }\label{tab:uncertainty} 
\centering
\vspace{0.1in}
\begin{tabular}{c|c} \hline\hline
  &  Syst. error (\%) \\
\hline
Tracking efficiency   &  14 \\
TOF matching & 10 \\
Electron selection & 2 \\
Hadron contamination & 0-20 \\
\hline
Sum of data uncertainties & 17-26 \\
\hline \hline

Particle yield & 8-24 \\
Branching ratio and form factors & 1-10 \\
\hline
Sum of simulation uncertainties & 11-27 \\

\hline \hline
\end{tabular}
\end{table}

%which described the CERES data~\cite{CERES} and the NA60 data~\cite{NA60} is also

\begin{figure}[htbp]
\begin{center}
\includegraphics[width=0.5\textwidth]{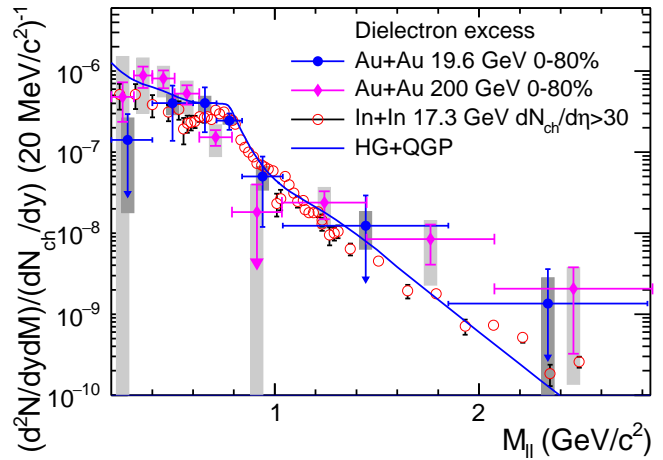}
\caption{(color online) The acceptance-corrected excess dielectron mass spectra, normalized to the charged particle multiplicity at mid-rapidity $dN_{ch}/dy$, in Au+Au collisions at $\sqrt{s_{NN}} = 19.6$ (solid circles) and 200 GeV (diamonds). The $dN_{ch}/dy$ values in Au+Au collisions at $\sqrt{s_{NN}} =$ 19.6 and 200 GeV are from Refs.~\cite{PID:15} and~\cite{PID:09}, respectively. Comparison to the NA60 data~\cite{NA60,NA60thermal} for In+In collisions at $\sqrt{s_{NN}} = $ 17.3 GeV (open circles) is also shown. Bars are statistical uncertainties, and systematic uncertainties are shown as grey boxes. A model calculation (solid curve)~\cite{ralf,ralfPrivate} with a broadened $\rho$ spectral function in hadron gas (HG) and QGP thermal radiation is compared with the excess in Au+Au collisions at $\sqrt{s_{NN}}$  = 19.6 GeV. The normalization uncertainty from the STAR measured $dN/dy$  is about 10\%, which is not shown in the figure. }\label{fig:excesscomp} 
\end{center}
\end{figure}

To quantify the yield, the known hadronic cocktail, $c\bar{c}\rightarrow e^{+}e^{-}X$ and Drell-Yan contributions were subtracted from the dielectron mass spectrum at $\sqrt{s_{NN}} = 19.6$ GeV. At $\sqrt{s_{NN}} = 200$ GeV, the known hadronic sources, $c\bar{c} \rightarrow e^{+}e^{-}X$, $b\bar{b} \rightarrow e^{+}e^{-}X$, and Drell-Yan contributions were subtracted. The excess dielectron mass spectra, corrected for detector acceptance, are shown in Fig.~\ref{fig:excesscomp} for Au+Au MB collisions at $\sqrt{s_{NN}} = 19.6$ and 200 GeV. The spectra are normalized to mid-rapidity $dN_{\rm ch}/dy$ in absolute terms to cancel out the volume effect, and compared with the excess dimuon yields from the NA60 measurements in In+In collisions at $\sqrt{s_{NN}} = 17.3$ GeV. The model calculation~\cite{ralf,ralfPrivate} including a broadened $\rho$ spectral function and QGP thermal radiation is consistent with the acceptance-corrected excess in Au+Au collisions at $\sqrt{s_{NN}} = $ 19.6 GeV. The excess at $\sqrt{s_{NN}} = $ 200 GeV  is higher than that at  $\sqrt{s_{NN}} = $ 17.3 GeV in the LMR and IMR, but within 2$\sigma$ uncertainty. Further measurements with better precision are needed to obtain the average temperature of the hot, dense medium created.

\begin{figure}[htbp]
\begin{center}
\includegraphics[width=0.5\textwidth]{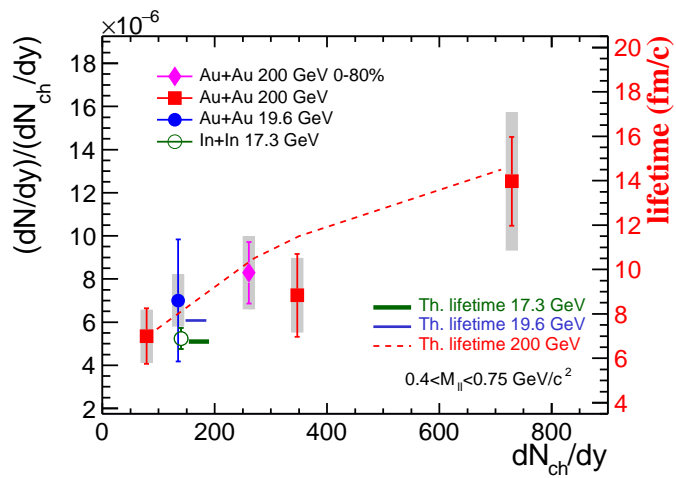}
\caption{(color online) Integrated yields of the normalized dilepton excesses for $0.4<M_{ll}<0.75$ GeV/$c^{2}$ as a function of $dN_{\rm ch}/dy$. The solid circle and diamond represent the results in 0-80\% Au+Au collisions at  $\sqrt{s_{NN}} = $ 19.6 and 200 GeV, respectively. The squares are the results for 40-80\%, 10-40\%, and 0-10\% Au+Au at $\sqrt{s_{NN}} = $ 200 GeV. The open circle represents the dimuon result from the NA60 measurement with $dN_{ch}/d\eta>30$. Bars are statistical uncertainties, and systematic uncertainties are shown as grey boxes. The theoretical lifetimes for $\sqrt{s_{NN}} = $ 200 GeV Au+Au as a function of $dN_{\rm ch}/dy$ in the model calculations~\cite{rapp:14} are shown as a dashed curve. The lifetimes for $\sqrt{s_{NN}} = $ 17.3 GeV In+In and $\sqrt{s_{NN}} = $ 19.6 GeV Au+Au in the same model calculations~\cite{rapp:14} are shown as the two horizontal bars. The $dN_{\rm ch}/dy$ values for the horizontal bars are shifted for clarity. }\label{fig:excessLMRcomp}
\end{center}
\end{figure}

Figure~\ref{fig:excesscomp} shows that the excess dielectron yield in the LMR at $\sqrt{s_{NN}} = $ 19.6 GeV has a magnitude similar to the excess dimuon yield at $\sqrt{s_{NN}} = $ 17.3 GeV. To quantitatively compare the excess in the LMR, the integrated excess yields of dielectrons in the mass region $0.4<M_{ll}<0.75$ GeV/$c^{2}$ are shown in Fig.~\ref{fig:excessLMRcomp} for 0-80\% Au+Au collisions at $\sqrt{s_{NN}} = $ 19.6 and 200 GeV. The results in finer centralities 0-10\%, 10-40\%, and 40-80\% are also shown for $\sqrt{s_{NN}} = $ 200 GeV collisions. The excess yield has a centrality dependence and increases from peripheral to central collisions at $\sqrt{s_{NN}} = $ 200 GeV. Comparing to the results from In+In collisions at $\sqrt{s_{NN}} = $ 17.3 GeV, the excess yield at $\sqrt{s_{NN}} = $ 19.6 GeV is consistent within the uncertainties while the excess at $\sqrt{s_{NN}} = $ 200 GeV is higher in central collisions, but within 2$\sigma$ uncertainty. This might indicate that the lifetime of the medium created in central collisions at $\sqrt{s_{NN}} = $ 200 GeV is longer than those in peripheral collisions and at $\sqrt{s_{NN}} = $ 17.3 GeV, which enhances the dilepton production from thermal radiation. 
The same model calculations~\cite{ralf,ralfPrivate}  that consistently describe the dilepton excesses in the $\sqrt{s_{NN}} = $ 17.3, 19.6, and 200 GeV A+A data give lifetimes of $6.8\pm1.0$ fm/\textit{c}, $7.7\pm1.5$ fm/\textit{c}, and $10.5\pm2.1$ fm/\textit{c}  for the 17.3 GeV In+In, 19.6 GeV Au+Au, and 200 GeV Au+Au data as shown in Fig.~\ref{fig:excessLMRcomp}~\cite{rapp:14}. In addition, the lifetime has a strong centrality dependence in $\sqrt{s_{NN}} = $ 200 GeV Au+Au collisions in the calculations, as indicated by the dashed curve in Fig.~\ref{fig:excessLMRcomp}. With the total baryon density nearly a constant and the dilepton emission rate dominant in the critical temperature region at $\sqrt{s_{NN}} = $  17.3-200 GeV, the normalized excess dilepton yields in the low mass region from the measurements are proportional to the calculated lifetimes of the medium~\cite{rapp:14}. We note that the lifetime might be model dependent. It is important to have the calculated lifetimes from other models to verify this proportionality.

\section{Summary}
\label{Summary}
In summary, the dielectron mass spectrum is measured in Au+Au collisions at  $\sqrt{s_{NN}} = $ 19.6 GeV by the STAR experiment at RHIC. Compared with known hadronic sources, a significant excess is observed, which can be consistently described in all beam energies by a model calculation in which a broadened $\rho$ spectral function scenario at low temperature and chiral symmetry restoration are included. Furthermore, the excess dielectron mass spectra, corrected for the STAR detector acceptance, are reported for the first time in Au+Au collisions at  $\sqrt{s_{NN}} = $ 19.6 and 200 GeV. In the LMR, the excess yield at $\sqrt{s_{NN}} = $ 19.6 GeV, normalized to the charged particle multiplicity $dN_{\rm ch}/dy$, is comparable to that in In+In collisions at $\sqrt{s_{NN}} = $ 17.3 GeV. For $\sqrt{s_{NN}} = $  200 GeV, the normalized excess yield is higher in central collisions than that at $\sqrt{s_{NN}} = $ 17.3 GeV and increases from peripheral to central collisions. These measurements indicate that the hot, dense medium created in central Au+Au collisions at top RHIC energy has a longer lifetime than those in peripheral collisions and at $\sqrt{s_{NN}} = $ 17.3 GeV.

\section{Acknowledgement}
\label{Acknowledgement}
We thank the RHIC Operations Group and RCF at BNL, the NERSC Center at LBNL, the KISTI Center in Korea, and the Open Science Grid consortium for providing resources and support. This work was supported in part by the Offices of NP and HEP within the U.S. DOE Office of Science, the U.S. NSF, CNRS/IN2P3, FAPESP CNPq of Brazil,  the Ministry of Education and Science of the Russian Federation, NNSFC, CAS, MoST and MoE of China, the Korean Research Foundation, GA and MSMT of the Czech Republic, FIAS of Germany, DAE, DST, and CSIR of India, the National Science Centre of Poland, National Research Foundation (NRF-2012004024), the Ministry of Science, Education and Sports of the Republic of Croatia, and RosAtom of Russia.



%% References


%%\bibliographystyle{model1-num-names}
%%\bibliography{<your-bib-database>}



\end{document}